\documentclass[twocolumn,showpacs,preprintnumbers,amsmath,amssymb]{revtex4}

\usepackage{graphicx}
\usepackage{dcolumn}
\usepackage{bm}

\begin{document}

\title{Origin of the Temperature Oscillation in Turbulent Thermal Convection}

\author{Heng-Dong Xi, Sheng-Qi Zhou, Quan Zhou, Tak-Shing Chan, and Ke-Qing Xia}
\address{Department of Physics, The Chinese University of Hong Kong,
Shatin, Hong Kong, China}

\date{\today}

\begin{abstract}
We report an experimental study of the three-dimensional spatial
structure of the low-frequency temperature oscillations in a
cylindrical Rayleigh-B\'{e}nard convection cell. Through
simultaneous multipoint temperature measurements it is found that,
contrary to the widely accepted scenario, thermal plumes are
emitted neither periodically nor alternatively, but randomly and
continuously, from the top and bottom plates. We further identify
a new flow mode---the sloshing mode of the large scale circulation
(LSC). This sloshing mode, together with the torsional mode of the
LSC, are found to be the origin of the oscillation of the
temperature field.
\end{abstract}


\maketitle

Thermal convection is a phenomenon that occurs widely in nature
and in many industrial processes \cite{kadanoff2001}. A paradigm
to study the generic convection phenomenon is the
Rayleigh-B\'{e}nard (RB) system \cite{ahlers2008rmp}.  A
fascinating feature of turbulent RB convection is the emergence of
a well-defined coherent oscillation in the presence of turbulent
background. This robust oscillation has been observed in both the
temperature \cite{castaing1989jfm} and velocity fields
\cite{qiu2000pre}, and in convection systems with different fluids
\cite{castaing1989jfm,sano1996prl,brown2007jsm} and different
geometries \cite{niemela2006jfm,thess2007pre,zhousq2007pre}.
Although much effort has been devoted to the study of this
phenomenon in the past
\cite{castaing1989jfm,villermaux1995prl,cioni1997jfm}, the nature
of this oscillation remains unsettled. It has been proposed that
this oscillation is due to the periodic emission of thermal plumes
from the top and bottom boundary layers of the system, which are
coupled by the large-scale circulation (LSC)
\cite{villermaux1995prl}. In this scenario, plume emission due to
boundary layer instability in one plate is triggered by the
arrival of thermal plumes from the other plate, which implies that
plumes are emitted not only periodically but also alternatively
between the top and bottom plates. Some later experimental studies
appear to support this picture and periodic plume emission has
since been attributed to be the source of the observed temperature
and velocity oscillations in the bulk fluid
\cite{ciliberto1996pre,qiu2001prl,sano2005prl,sun2005prea}. In
addition to the apparent oscillations observed for temperature and
velocity in the vertical circulation plane of the LSC, horizontal
oscillations of the velocity field have also been observed
\cite{funfschilling2004prl,xi2006pre,resagk2006pof} and it has
been suggested that periodic plume emission is not necessary for
the horizontal oscillation of the bulk fluid
\cite{funfschilling2004prl,resagk2006pof}. Recently, it is
conjectured that the local temperature oscillations may be caused
by the torsional motion of the LSC \cite{ahlers2008rmp}. The
conjecture, which could explain the oscillation near the top and
bottom plate, is unable to explain that observed at the mid-height
plane, since, by symmetry, the torsional oscillations cancel out
at the mid-height plane. We also note that the experimental
studies that appear to show evidence of periodic plume emissions
are two-dimensional (2D) measurements
\cite{ciliberto1996pre,qiu2001prl,sano2005prl,sun2005prea}. With
the LSC's twisting oscillation near the top and bottom plates and
its azimuthal meandering
\cite{funfschilling2004prl,xi2006pre,brown2006jfm},
three-dimensional (3D) measurements become essential to unlock the
intricate flow dynamics.

\begin{figure}[h]
\begin{center}
\resizebox{1\columnwidth}{!}{%
\includegraphics{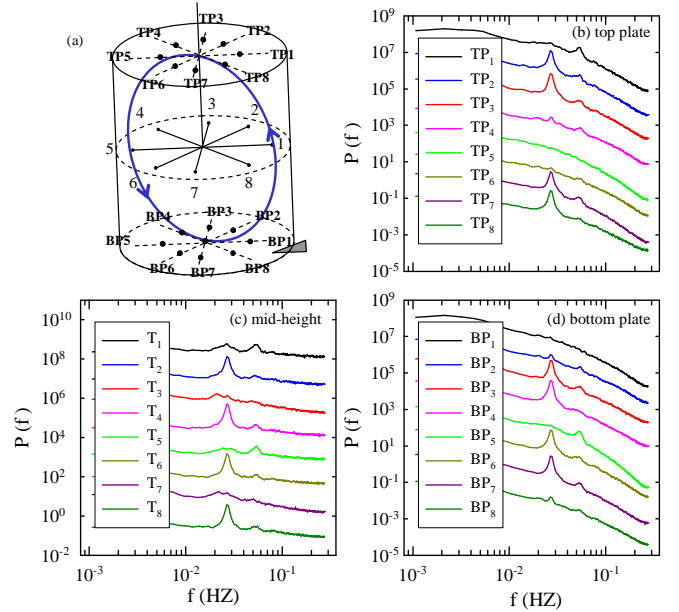}}
\caption{\label{setup} (Color online) (a) Sketch of the convection
cell and thermistor configurations. Power spectra of temperatures
measured by thermistors (b) embedded in the top plate at positions
TP1 to TP8 (from top to bottom);  (c) at mid-height positions T1
to T8 in the bulk fluid; and (d) embedded in the bottom plate at
positions BP1 to BP8. For clarity, each data set is shifted up
from its lower neighbor by a factor of 20. }
\end{center}
\end{figure}

In this Letter, we present 3D measurements of the temperature
oscillations in a cylindrical convection cell of aspect ratio
unity. Local temperatures are measured simultaneously by 24 probes
placed inside the convecting fluid and embedded in the top and
bottom plates. These measurements reveal the origin of the
temperature oscillation in the system and shed light on the
driving mechanism of its flow dynamics. The experiment was
conducted in a RB convection cell that has been described
elsewhere \cite{sun2005prea}. Briefly it is a vertical cylinder of
height H = 19.0 cm and diameter D = 19.0 cm, with upper and lower
copper plates and Plexiglas sidewall. Water was used as the
working fluid and measurements were made at the Rayleigh number Ra
= 1.7 $\times 10^9$ and 5.0 $\times 10^9$, with Prandtl number Pr
= 5.3. As the two measurements give the same qualitative results,
only results for Ra = 5.0 $\times 10^9$ will be presented. Local
temperatures in the fluid are measured by 8 thermistors of 300
$\mu$m in diameter. As shown in Fig. \ref{setup}(a), these
thermistors are mounted on a star-shaped frame made of $1$ mm
diameter stainless steel tube. They have an equal azimuthal
separation and have a distance of 1 cm from the sidewall, where
the main ascending and descending flows are passing through
\cite{sun2005prea}. The frame is soldered perpendicularly to a $1$
mm diameter stainless steel tube and can thus traverse vertically
along the central axis of the cell. The thermistors are labelled
as 1, 2, ... 8, which also represent their azimuthal positions.
The temperatures in the top and bottom plates were measured by 16
thermistors of $2.5$ mm in diameter [shown as solid circles in
Fig. \ref{setup} (a)]. Eight of them are embedded in the top plate
at half radius from the plate center and about $2$ mm away from
the fluid contact interface at positions TP1 to TP8; while the
other eight are similarly embedded in the bottom plate at
positions BP1 to BP8. To lock the azimuthal orientation of the LSC
steadily, we tilted the convection cell with $2^\circ$ at position
1. Temperatures of the 8 small thermistors and 16 large
thermistors are measured simultaneously by a $6\frac{1}{2}$-digit
multimeter at a data rate of 0.55$/$sec.

Figures \ref{setup}(b) and (d) plot the frequency power spectra of
the temperatures measured by the embedded thermistors in the top
and bottom plates, respectively. Due to the finite thermal
diffusivity of the plates, when a hot (cold) plume is emitted from
the bottom (top) plate, it leaves a cold (hot) spot there for a
finite time before the temperature recovers to the previous value
by conduction. Similarly, when a hot (cold) plume impinge the top
(bottom) plate, it will heat (cool) fluid in the boundary layer
and leave a thermal imprint there. It has been shown in previous
experiments that thermistors embedded in the plates are able to
capture signatures of these plume departures and arrivals
\cite{cioni1997jfm,sun2007jfm}. The prominent peaks in the power
spectra near $f_0 (\simeq 0.028$ Hz) correspond to the same
oscillation frequency (for the same Ra) measured in many previous
studies both in the fluid and in the plates
\cite{castaing1989jfm,sano1996prl,ciliberto1996pre,cioni1997jfm,qiu2001prl}.
It is seen from the figures that strong oscillations are observed
at positions TP2, TP3, TP7, and TP8, where the hot ascending flow
arrives at the top plate, and at positions BP3, BP4, BP6, and BP7,
where the cold descending flow arrives at the bottom plate. On the
other hand, no significant oscillation is observed at positions
TP4, TP5, and TP6, where cold plumes are emitted, and at positions
BP1, BP2, and BP8, where hot plumes are emitted. These results
show clearly that there is no temperature oscillation at positions
where plumes are emitted. Furthermore, if the plumes are not
emitted periodically, then the oscillations observed at positions
where plumes arrive from the opposite plate must originate in the
bulk.

%
%
%

\begin{figure}[h]
\begin{center}
\resizebox{0.9\columnwidth}{!}{%
\includegraphics{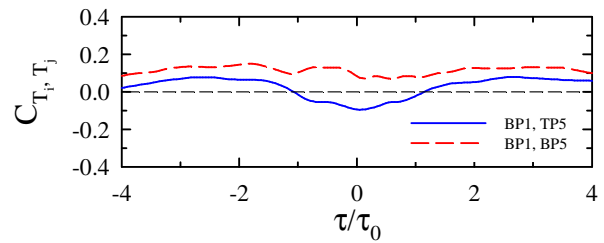}}
\caption{\label{corr_plate} (Color online)  Temperature
cross-correlation functions measured in-plate at where plumes
depart or arrive.}
\end{center}
\end{figure}

\begin{figure}[h]
\begin{center}
\resizebox
{0.9\columnwidth}{!}{%
\includegraphics{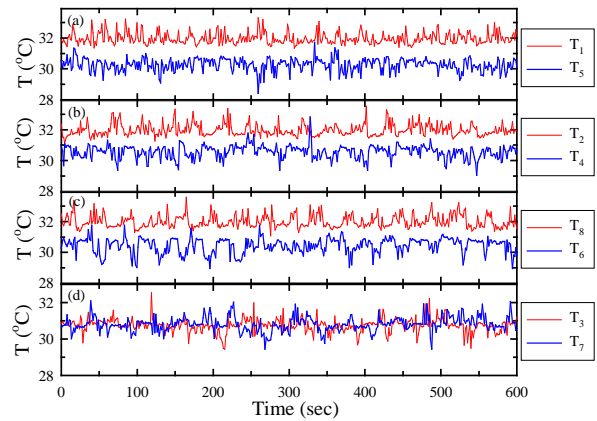}}
\caption{\label{trace4}  (Color online) Segments of time traces of
the temperature measured at the mid-height plane by thermistors
$T_1$ to $T_8$. For clarity, $T_1$, $T_2$ and $T_8$ are shifted up
by 1$^\circ$C.}
\end{center}
\end{figure}


An important feature of the models that attribute the origin of
temperature oscillations to the periodic emission of thermal
plumes from the boundary layer is that emission of the hot (cold)
plumes is triggered by arrival of cold (hot) plumes and thus hot
and cold plumes are emitted alternatively from the bottom and top
plates with a time separation equals to half of the oscillation
period. To test this, we examine the phase relationships between
the temperatures measured in the plates, which can be obtained
from the cross-correlation function of the measured temperatures:
$C_{T_i, T_j}(\tau) = \langle{(T_i(t+\tau)-\langle
T_i\rangle)(T_j(t)-{\langle{T_j}\rangle})\rangle}/{\sigma_{
{T_i}}\sigma_{T_j}}$, where $\sigma_{T_i}$ and $\sigma_{T_j}$ are
the standard deviations of the two quantities respectively. Figure
\ref{corr_plate} shows the cross-correlation function between
temperatures measured at the positions BP1 and TP5, where hot and
cold plumes are emitted respectively. If the hot and cold plumes
are shed alternatively from the bottom and the top plates, the
cross-correlation function $C_{BP1,TP5}(\tau)$ should have strong
negative peaks at $\tau = \pm\tau_0/2$, with $\tau_0/2$ the
approximate time for plumes to traverse the height of the cell
(here $\tau_0 = 1/f_0 = 36$ sec). As shown in the figure the
correlation is very weak and there is no negative peaks at $\tau =
\pm\tau_0/2$, but a negative peak at $\tau = 0$. A negative peak
at $\tau = 0$ means that the hot and cold plumes are shed
simultaneously. Also plotted in Fig. \ref{corr_plate} is the
cross-correlation function between temperatures measured at BP1
and BP5, the positions where in the bottom plate hot plumes depart
and cold ones arrive, respectively. If the arrival of the cold
plumes at BP5 triggers the emission of the hot plumes at position
BP1, the cross-correlation function will have a strong positive
peak near $\tau = 0$, which is clearly not the case as shown in
the figure.

\begin{figure}[h]
\begin{center}
\resizebox{0.9\columnwidth}{!}{%
\includegraphics{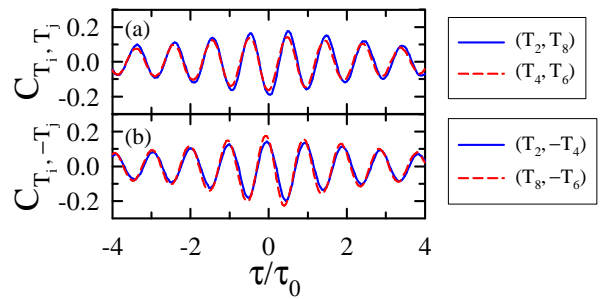}}
\caption{\label{correlation}  (Color online)  Temperature
cross-correlation functions measured in-fluid at mid-height
plane.}
\end{center}
\end{figure}

Figure \ref{setup}(c) shows the power spectra of the individual
temperatures $T_1$ to $T_8$ measured in the bulk fluid at the
mid-height (H/2) plane by the 8 small thermistors. The prominent
peaks in the power spectra have the same frequency $f_0$ as in
Figs. \ref{setup}(b) and (d). It is seen that no significant
oscillations are observed at positions 1 and 5, which are within
the circulation plane of the LSC.  On the other hand, strong
oscillations are present at positions 2, 4, 6 and 8. To understand
why oscillation is present at some positions while absent in
others at the mid-height plane, we plot in Fig. \ref{trace4}
segments of time traces of $T_1$ to $T_8$. It can be seen that at
positions 1, 2, 8 positive spikes dominate, suggesting hot plumes
are ascending at these positions. At positions 4, 5 and 6 negative
spikes dominate, suggesting cold plumes are descending there. At
positions 3 and 7, both hot plumes and cold plumes are present.
These temperature signals are consistent with a coherent large
scale circulatory flow with a band of about half a diameter wide.
To examine the phase relationships between the 8 temperature
signals, we study their cross-correlation functions. Figure
\ref{correlation}(a) shows the cross-correlations between the
$(T_2, T_8)$ pair and between the $(T_4, T_6)$ pair. The peak of
$C_{T_2, T_8}$ ($C_{T_4, T_6}$) near $\tau = \tau_0/2$ indicates
that the hot (cold) fluids pass the positions $2$ and $8$ ($4$ and
$6$ ) alternatively with a time delay of $\tau_0/2$. We calculate
$C_{T_i, -T_j}$ for a pair of thermistors when one of them mainly
senses the upward spikes and the other mainly downward spikes.
Figure \ref{correlation}(b) plots $C_{T_2, -T_4}$ and $C_{T_8,
-T_6}$ and the peak near $\tau = 0$ indicates that the hot fluids
pass $2$ and cold ones pass $4$ simultaneously. Similar
relationship exists between positions $6$ and 8. Simultaneous
presence of hot bursts at position 2 (8) and cold bursts at
position 4 (6) is another evidence that hot and cold plumes are
not emitted alternatively. The phase relationships above could
also be observed from the temperature time traces shown in Fig.
\ref{trace4}. Together, these results suggest a horizontal
oscillation of the bulk fluid along the direction perpendicular to
the vertical plane containing positions 1 and 5. In this
off-center or sloshing mode of the bulk fluid, hot ascending fluid
oscillates between positions 2 and 8 and cold descending fluid
oscillates between positions 4 and 6, and this explains why the
oscillation strength at these positions are approximately the same
[Fig. 1(c)]. It should be noted that signatures of this sloshing
mode has been observed by Qiu \emph{et al.} \cite{qiu2004pof} in a
previous local velocity measurements, who found that the strongest
oscillation occurs in the direction perpendicular to the LSC
plane.

\begin{figure}[h]
\begin{center}
\resizebox{0.9\columnwidth}{!}{%
\includegraphics{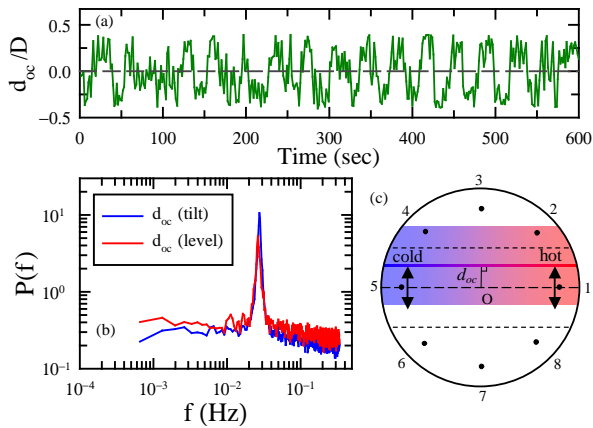}}
\caption{\label{doc}  (Color online) (a) Segment of time trace of
off-center distance $d_{oc}$. (b) Power spectra of $d_{oc}$ for
both levelled and tilted cells. (c) A schematic drawing of the
off-center motion of the LSC at mid-height plane. The shaded band
represents an instantaneous position of the horizontal
cross-section of the LSC with its center at a distance $d_{oc}$
away from its average position (long-dashed line), and the two
short-dashed lines represent the average position of the LSC band.
The black dots represent the actual positions of the 8 small
thermistors.}
\end{center}
\end{figure}

To study the sloshing oscillation of the bulk fluid
quantitatively, we determine the hottest and coldest azimuthal
positions of the bulk fluid along the sidewall from the
instantaneous azimuthal temperature profile measured by the 8
thermistors at the mid-height plane. These positions are
determined by making a quadratic fit around the hottest and
coldest temperature readings respectively \cite{tobesubmitted}.
The line connecting these two positions is the central line of the
LSC band. The orientation of this line is the orientation of the
LSC's vertical circulating plane, which is found to be very
similar to that obtained by fitting a cosine function to the
instantaneous azimuthal temperature profile \cite{tobesubmitted}.
We define the distance between this line and the cell's central
vertical axis as the off-center distance $d_{oc}$, which exhibits
a periodic oscillation with an amplitude about $1/3$ of the cell
diameter [Fig. \ref{doc}(a)]. Figure \ref{doc}(b) shows the power
spectrum of $d_{oc}$, which has a peak at $f_0$ ($0.028$ Hz). Also
plotted in the figure is the power spectrum of $d_{oc}$ when the
cell is levelled. It shows that the sloshing mode exists in both
tilted and levelled cases and that it is not a result of the LSC
being locked in a fixed azimuthal orientation. Figure \ref{doc}(c)
draws schematically this sloshing mode of the bulk fluid at
mid-height plane based on the measured properties of $d_{oc}(t)$.
This picture confirms some of the above conclusions inferred from
the properties of cross-correlation functions between the local
temperature probes. For example, it is seen that due to this
periodical sloshing motion of the LSC, the thermistors at
positions 2 and 8 (positions 4 and 6) will alternatively
experience the hot ascending (cold descending) flow. For the
positions 1 and 5, as the LSC has a width of roughly half cell
diameter, they will always experience the hot ascending and cold
descending flow but the degree of `hot' and `cold' varies due to
the off-center oscillation. This explains the very weak peak at
$f_0$ in the power spectra of $T_1$ and $T_5$. The peak at 2$f_0$
of these spectra is due to the fact that within one period of the
off-center motion the central line of the LSC crosses the (1,5)
plane twice. From Fig. 5 (c) it is seen that the positions 3 and 7
are farthest from the band of LSC and so they can barely sense the
bulk off-center oscillation. This is evidenced by the barely
visible peaks at $f_0$ in the power spectra of $T_3$ and $T_7$.

Some of the features observed by the thermistors embedded in
plates can also be understood now. The oscillation with frequency
$f_0$ at positions TP2, TP3, TP7, TP8, BP3, BP4, BP6 and BP7, and
the oscillation with frequency $2f_0$ at positions TP1 and BP5 are
due to the the sloshing plus twisting motion of LSC. On the other
hand, no significant oscillation is seen at positions BP1, BP2,
BP8, TP4, TP5 and TP6. But when we placed the 8 small thermistors
very near (3 mm) the bottom plates, positions 2 and 8 have strong
oscillations. This is because that the horizontal positions of the
plumes are modulated by the flow field only \emph{after} they
leave the respective plates and are thus not sensed by the
embedded probes. This is a further evidence that the temperature
oscillation does not originate from the boundary layers. It may be
noted that some recent models of the LSC dynammics
\cite{resagk2006pof,brown2007prl} have ignored the thermal
boundary layer and the plumes. Thus, these models could in
principle be modified to explain the results reported here.

With the above picture, we can understand the reason why some
earlier measurements made in a 2D plane appear to observe that
thermal plumes are emitted periodically. In the present experiment
the orientation of the LSC is locked steadily in the $(1,5)$ plane
by tilting the cell with a $2 ^\circ$ angle so that it has a very
small range of azimuthal angular fluctuation ($\phi_{rms} = $
8.5$^\circ$). In some of the previous studies that observed
temperature oscillations within the plane of the LSC, the
convection cell was titled by lass than 1$^\circ$
\cite{qiu2001prl,qiu2004pof,sun2005prea}. When we tilted our cell
by $0.5^\circ$ it is found that the LSC is able to explore much
broader azimuthal range ($\phi_{rms} = $ 22.6$^\circ$)
\cite{ahlers2006jfm}. When this azimuthal meandering is combined
with the sloshing motion the LSC band could leave and return to
the (1,5) plane so that the alternative occurrence of the hot and
cold bursts in $T_1$ and $T_5$ can be observed. In fact in this
case temperature oscillations at all 8 positions can be observed
\cite{tobesubmitted}. Therefore the alternate appearance of hot
and cold bursts at positions of 1 and 5 when the azimuthal
orientation of the LSC is not steadily locked can be understood as
a result of the horizontal motion of hot ascending and cold
descending fluids being modulated by the sloshing mode. To
conclude, our simultaneous multipoint temperature measurements
show directly and convincingly that thermal plumes are emitted
neither periodically nor alternatively from the top and bottom
plates and that temperature oscillations are caused by the
sloshing mode and the torsional mode of the velocity field in the
central and boundary layer regions of the system, respectively.

We gratefully acknowledge support of this work by the Hong Kong
Research Grants Council under Grant Nos. CUHK 403806 and 404307.


\end{document}